\documentclass[aps,prd,reprint,showkeys,showpacs,superscriptaddress]{revtex4-1}

\usepackage{amssymb}
\usepackage{amsmath}
\usepackage{amsfonts}
\usepackage{bm}
\usepackage{graphicx}
\usepackage{mathtools}
\usepackage{here}
\usepackage{xcolor}

\usepackage[colorlinks=true]{hyperref}

\usepackage{braket}
\begin{document}

\newcommand{\sub}[1]{\ensuremath{_{\mathrm{#1}}}}
\newcommand{\sps}[1]{\ensuremath{^{\mathrm{#1}}}}

\title{Duality in the dynamics of Unruh-DeWitt detectors in conformally related spacetimes
}

\author{Masahiro Hotta}
\affiliation{Graduate School of Science, Tohoku University,Sendai, 980-8578, Japan}

\author{Achim Kempf}
\affiliation{Department of Applied Mathematics, University of Waterloo, Waterloo, Ontario, N2L 3G1, Canada}
\affiliation{Institute for Quantum Computing, University of Waterloo, Waterloo, Ontario, N2L 3G1, Canada}
\affiliation{Perimeter Institute for Theoretical Physics, Waterloo, Ontario N2L 2Y5, Canada}
\affiliation{Department of Physics and Astronomy, University of Waterloo, Waterloo, Ontario, N2L 3G1, Canada}

\author{Eduardo Mart\'in-Mart\'inez}
\affiliation{Department of Applied Mathematics, University of Waterloo, Waterloo, Ontario, N2L 3G1, Canada}
\affiliation{Institute for Quantum Computing, University of Waterloo, Waterloo, Ontario, N2L 3G1, Canada}
\affiliation{Perimeter Institute for Theoretical Physics, Waterloo, Ontario N2L 2Y5, Canada}

\author{Takeshi Tomitsuka}
\affiliation{Graduate School of Science, Tohoku University,Sendai, 980-8578, Japan}

\author{Koji Yamaguchi}
\affiliation{Graduate School of Science, Tohoku University,Sendai, 980-8578, Japan}

\begin{abstract}
We prove a nonperturbative duality concerning the dynamics of harmonic-oscillator-type Unruh-DeWitt detectors in curved spacetimes.  
Concretely, using the Takagi transformation 
we show that the action of a harmonic oscillator Unruh-DeWitt detector with one frequency in a spacetime is equal to that of a detector with a different frequency in a conformally related spacetime.
As an example, we show that the dynamics of simple stationary detectors in flat spacetime is dual to that of detectors in a cosmological scenario. 
The nonperturbative duality enables us to investigate entanglement harvesting in new scenarios in curved spacetime by using results obtained in simpler, conformally related spacetimes.
\end{abstract}
\maketitle

\section{Introduction}

It has been long known that the vacuum state of a quantum field theory (QFT) can contain quantum correlations between spacelike separated regions \cite{Alegbra1,Alegbra2}. However, it is not trivial to understand these correlations in terms of experimentally accessible measurements. One of the reasons for this is that the formal treatment of measurements in QFT is much more complex than its already challenging nonrelativistic counterpart. For example, projective measurements are problematic in QFT even in the most trivial cases \cite{Sorkin1993_proceeding}. Moreover, even the standard notion of particle as a localized system ---oftentimes used to give intuition about observable measurements in a QFT--- has been shown to be fundamentally not compatible with the formulation of a consistent relativistic quantum theory \cite{Malament1,Malament2,Malament3}. 

The challenge of performing measurements on a quantum field has been tackled with the introduction of idealized models of particle detectors \cite{Takagi,DeWitt,Crispino,Unruh-Wald} defined as nonrelativistic quantum systems that couple locally to quantum fields. A key utility of such detector systems is that they provide an operational way to obtain local information about the quantum field. Further, the excitation of such a detector system can be interpreted as the localized absorption of a field quantum. In this way, these detector models have helped redefine the elusive notion of particle in QFT.  

Particle detectors have proven to be versatile tools in the study of QFT in curved spacetimes, see e.g.,   \cite{Crispino,Takagi} in that they, for example, played a fundamental role in the  the operational formulation of the Hawking and Unruh effects (see, e.g., \cite{unruh_notes_1976,Unruh-Wald,candelas_irreversible_1977,Carballo2019}). Besides their many uses in fundamental QFT, particle detectors are also extensively used to model experimental setups in quantum optics \cite{scullybook} and in superconducting circuits \cite{Wallraff:2004aa}. Indeed, an atom dipolarly coupled to the electromagnetic field can accurately be described with a particle detector model for the second-quantized electromagnetic field, capturing the fundamental features of the interaction of matter and light \cite{Pozas2016,MMQO2018,scullybook}. Moreover, In relativistic quantum information, particle detectors have also seen many uses, for example in modeling classical and quantum communication through quantum fields \cite{ClicheKempfD,Jonsson1,Jonsson2,Jonsson3,Jonsson4,MMQO2018,Katja,Landulfo,Funai,Shockwaves,Petar}.

The most common particle detector model, the Unruh-DeWitt (UDW) model~\cite{unruh_notes_1976,DeWitts}, can be found in the literature in several slightly different (but fundamentally similar) formats, e.g., a field in a box \cite{unruh_notes_1976}, a two-level system \cite{DeWitts} or a harmonic oscillator (see, e.g., \cite{BeiLok,Brown2012,Fuenetesevolution}). It consists of a localized linear coupling of an internal degree of freedom of the detector with a scalar field. 

UDW detectors have been particularly useful for the study of entanglement harvesting. Entanglement harvesting is the generic name for processes where two or more particle detectors get entangled through their interaction with the field even when remaining spacelike separated. In this case, the detectors are harvesting the spacelike entanglement contained in the ground state and excited states of quantum fields that we mentioned in the beginning of this introduction \cite{Alegbra1,Alegbra2}. Entanglement harvesting has been extensively studied in different contexts from fundamental and applied perspectives in both flat spacetimes (e.g., \cite{Valentini1991,Reznik2003,Reznik1,reznik2,Pozas-Kerstjens:2015,Nick,Olson2011,Olson2012,Salton:2014jaa,Pozas2016,Nambu2013,Kukita2017,Kukita20172,Ng2,Henderson2018,Henderson2019,PetarHarv,Cong2019,Brown1,Brown2,Retzker2005,Sabinprl,Farming}) as well as in a some curved spacetimes, such as de Sitter spacetime \cite{Nambu2013,Kukita2017,Kukita20172}, anti-de Sitter spacetime \cite{Ng2,Henderson2019} and the BTZ spacetime \cite{Henderson2018}.

Computing particle detector responses exactly is challenging, however, apart from a handful of analytically accessible scenarios considered in the literature. Even flat spacetime calculations can quickly become very complex when the detector trajectories are noninertial. The situation is even more challenging when considering entanglement harvesting, where the calculation of measures of entanglement generation are technically more involved than the calculation of transition probabilities or transition rates of particle detectors.

In this paper we develop a new tool that can map a class of  difficult problems of the dynamics of particle detectors in curved spacetimes to solvable, simpler problems in flat spacetimes through the use of the Takagi transformation \cite{Takagitfm1,Takagitfm2}. This tool will be particularly helpful in problems involving several particle detectors coupling to quantum fields, as for example scenarios of entanglement harvesting.


Concretely, in this paper, we show a duality in the dynamics of harmonic-oscillator-type UDW detectors in curved spacetimes. 
The action of a harmonic oscillator can be recast into the action of a free particle through the Takagi transformation \cite{Takagitfm1,Takagitfm2}, which is composed of a nonlinear transformation of time and a linear transformation of the position operator. In general, this duality connects the dynamics of harmonic oscillators with different angular frequencies. Under the Takagi transformation, an affine time parameter is nonlinearly mapped into another time parameter. We introduce a conformal factor so that the transformed time becomes affine in a new spacetime conformally related to the original one. For an UDW detector linearly coupled to the field, the transformation of the position operator can be interpreted as a modification of switching and smearing functions which characterize the temporal and spatial distributions of the interaction. By using these facts, we prove the duality in dynamics of harmonic oscillator UDW detector systems in different spacetimes related through the conformal-Takagi transformation. Since the conformal-Takagi transformation is parameterized by a non-negative number, there are infinitely many detector systems in different spacetimes whose dynamics is equivalent. 

The duality can also be extended to the case where multiple UDW detectors are coupled to a scalar field. As an example, we investigate entanglement harvesting protocols. By using the duality, it is shown that the harmonic oscillator UDW detector systems related through the conformal-Takagi transformation extract the same amount of entanglement with each other. 
This nonperturbative duality enables us to investigate the entanglement structure of the field in a curved spacetime by using a result in conformally related spacetimes. 

We further show that the difference in the `negativity' measure of entanglement between qubit UDW detectors and harmonic oscillator UDW detectors only appears in the subleading terms. This means that any leading-order calculation of the negativity using qubit detectors which interact with a conformally coupled scalar field can be mapped into the corresponding calculations in different conformally related spacetimes.

Throughout this manuscript, we regard spacetimes as background, neglecting the backreaction from fields and detectors.
We take the signature of metric as $(-,+,\cdots,+)$. Furthermore, we  adopt  natural units $c=\hbar=1$ for our calculations.

\section{Nonperturbative Duality in dynamics of harmonic oscillator UDW detectors}\label{sec_dynamics}
In this section, we investigate a harmonic oscillator UDW detector system with a conformally coupled massless scalar field. We show a duality among the dynamics in conformally related spacetimes. We first focus on pointlike detectors for simplicity. The result can be extended to smeared detectors. 

\subsection{Takagi transformation}
Let us first review the Takagi transformation \cite{Takagitfm1,Takagitfm2}, which reveals a relation in dynamics between two harmonic oscillators with different angular frequencies. 
The action for a harmonic oscillator with angular frequency $\omega$ and unit mass is given by
\begin{align}
 S=\frac{1}{2}\int \mathrm{d}\lambda \left(\left(\frac{\mathrm{d}q_{\omega}(\lambda)}{\mathrm{d}\lambda}\right)^2-\omega^2 q_{\omega}^2\right),
\end{align}
where $\lambda$ denotes the proper time. Introducing a new time coordinate
\begin{align}
 \tau(\lambda)&\coloneqq \frac{1}{\Omega}\arctan{\left(\frac{\Omega}{\omega}\tan{(\omega \lambda)}\right)},\label{eq_Takagi1}
\end{align}
and a new variable
\begin{align}
 q_{\Omega}(\tau)\coloneqq \frac{\cos{(\Omega\tau)}}{\cos{(\omega\lambda(\tau))}}q_{\omega}(\lambda(\tau))\label{eq_Takagi2}
\end{align}
with a positive parameter $\Omega$, a straightforward calculation shows that the action can be recast in
\begin{align}
 S=\frac{1}{2}\int \mathrm{d}\tau \left( \left(\frac{\mathrm{d}q_{\Omega}(\tau)}{\mathrm{d}\tau}\right)^2-\Omega^2 q_{\Omega}^2\right),
\end{align}
which is the action for a harmonic oscillator with angular frequency $\Omega$. 
The transformation defined by Eqs.\eqref{eq_Takagi1} and \eqref{eq_Takagi2} is called the Takagi transformation. It should be noted that the new parameter $\tau$ is not a proper time since
\begin{align}
 \frac{\mathrm{d}\tau(\lambda)}{\mathrm{d}\lambda}=\frac{1}{\cos^2{(\omega\lambda)}+\left(\frac{\Omega}{\omega}\right)^2\sin^2{(\omega\lambda)}}\neq 1\label{eq_dtaudlambda}
\end{align}
holds for $\Omega\neq \omega$.

For quantized harmonic oscillators, the Takagi transformation corresponds to a unitary transformation. Consider a harmonic oscillator whose Hamiltonian is given by $\hat{H}_\omega=\frac{\omega^2}{2}\hat{q}^2+\frac{1}{2}\hat{p}^2$ that generates the time evolution with respect to a time parameter $\lambda$. For $\tau$ defined in Eq.\eqref{eq_Takagi1}, it holds that
\begin{align}
 \hat{V}_{\omega}(\lambda)\hat{U}_{\omega}(\lambda)=\hat{V}_{\Omega}(\tau(\lambda))\hat{U}_{\Omega}(\tau(\lambda)),\label{eq_takagi_unitary}
\end{align}
where we have defined
\begin{align}
\hat{V}_{\omega}(\lambda)&\coloneqq e^{i\frac{\ln{\left(\cos(\omega\lambda)\right)}}{2}(\hat{q}\hat{p}+\hat{p}\hat{q})}e^{i\frac{\tan{(\omega\lambda)}}{2}\hat{q}^2},\\ \hat{U}_{\omega}(\lambda)&\coloneqq e^{-i\left(\frac{\omega^2}{2}\hat{q}^2+\frac{1}{2}\hat{p}^2\right)\lambda}, \\
 \hat{V}_{\Omega}(\tau)&\coloneqq e^{i\frac{\ln{\left(\cos(\Omega\tau)\right)}}{2}(\hat{q}\hat{p}+\hat{p}\hat{q})}e^{i\frac{\tan{(\Omega\tau)}}{2}\hat{q}^2},\\ \hat{U}_{\Omega}(\tau)&\coloneqq e^{-i\left(\frac{\Omega^2}{2}\hat{q}^2+\frac{1}{2}\hat{p}^2\right)\tau} .
\end{align}
See the Appendix for the proof. In particular, it is shown from Eq.\eqref{eq_takagi_unitary} that
\begin{align}
 \hat{q}_{\Omega}(\tau(\lambda))=\frac{\cos{(\Omega\tau(\lambda))}}{\cos{(\omega\lambda)}}\hat{q}_{\omega}(\lambda)\label{eq_takagi_q}
\end{align}
holds, where we have defined
\begin{align}
  \hat{q}_{\Omega}(\tau)\coloneqq \hat{U}_{\Omega}(\tau)^\dag\hat{q}\hat{U}_{\Omega}(\tau),\quad  \hat{q}_{\omega}(\lambda)\coloneqq \hat{U}_{\omega}(\lambda)^\dag\hat{q}\hat{U}_{\omega}(\lambda).
\end{align}
Note that the expression in Eq.\eqref{eq_Takagi1} is valid only in an interval $\lambda\in(-\pi/(2\omega),\pi/(2\omega))$. This does not mean that the Takagi transformation is applicable only for a short time interval. For example, let us define $\tau(\lambda)$ as
\begin{align*}
 &\Omega\tau(\lambda)=\arctan{\left(\frac{\Omega}{\omega}\tan{\left(\omega\lambda-\pi n\right)}\right)}+\pi n\\
 &\left(\text{for} -\frac{\pi}{2}+\pi n\leq \omega\lambda < \frac{\pi}{2}+\pi n,\quad n\in\mathbb{Z}\right)
\end{align*}
for $\Omega>0$.
Then, $\tau(\lambda)$ is a one-to-one map from $\mathbb{R}$ to itself. For this function $\tau(\lambda)$, Eq.\eqref{eq_takagi_q} holds. It should be noted that $\frac{\mathrm{d}\tau(\lambda)}{\mathrm{d}\lambda}$ is given by Eq.\eqref{eq_dtaudlambda} for any $\lambda$. 

\subsection{Conformal-Takagi transformation}
We consider a pointlike UDW detector in an $(n+1)$-dimensional spacetime $(M,g_{\mu\nu})$ whose internal degrees of freedom are modeled by a harmonic oscillator. When it interacts with a conformally coupled massless scalar field, the action for the system is given by $S_{\mathrm{UDW}}=S_{\mathrm{free},\phi}+S_{\mathrm{free},\omega}+S_{\mathrm{int}}$, where
\begin{align}
 &S_{\mathrm{free},\phi}\coloneqq-\frac{1}{2}\int_M \mathrm{d}^{n+1}x \sqrt{-g}\left(g^{\mu\nu}\nabla_\mu\phi(x) \nabla_\nu \phi(x) \right.\nonumber\\
 &\left.\quad\quad\quad\quad\quad\quad\quad\quad\quad\quad\quad\quad\quad\quad+\xi R(x)\phi(x)^2 \right),\label{eq_def_fieldaction}\\
&S_{\mathrm{free},\omega}\coloneqq \frac{1}{2}\int \mathrm{d}\lambda \left(\left(\frac{\mathrm{d}q_{\omega}(\lambda)}{\mathrm{d}\lambda}\right)^2-\omega^2 q_{\omega}^2(\lambda)\right),\\
&S_{\mathrm{int}}\coloneqq -  c \int  \mathrm{d}\lambda\sqrt{2\omega}\chi_{\omega}(\lambda)q_{\omega}(\lambda)\phi(x^{\mu}(\lambda)),
\end{align}
where $\nabla_\mu$ denotes the covariant derivative associated with $g_{\mu\nu}$ and $\xi=\frac{n-1}{4n}$ characterizes a coupling between the field and the Ricci scalar $R(x)$ determined by the metric $g_{\mu\nu}$. The factor $\sqrt{2\omega}$ in the interaction term is included for future convenience. In the interaction term, $c$ is the coupling constant, $x^{\mu}(\lambda)$ denotes the trajectory of the detector with an affine parameter $\lambda$, and $\chi_{\omega}(\lambda)$ is the switching function which characterizes the temporal dependence of the interaction between the detector and the field.

As is seen in the previous section, under the Takagi transformation:
\begin{align}
 \lambda\to\tau(\lambda),\quad q_{\omega}\to q_{\Omega}(\tau)\coloneqq \frac{\cos{(\Omega\tau)}}{\cos{(\omega\lambda(\tau))}}q_{\omega}(\lambda(\tau)),
\end{align}
the free part for the harmonic oscillator system $S_{\mathrm{free},\omega}$ is rewritten as
\begin{align}
 S_{\mathrm{free},\omega}=\int \mathrm{d}\tau \left( \left(\frac{\mathrm{d}q_{\Omega}(\tau)}{\mathrm{d}\tau}\right)^2-\Omega^2 q_{\Omega}^2(\tau)\right).
\end{align}
However, $\tau$ is not a proper time for the trajectory in $(M,g_{\mu\nu})$ as is mentioned before. 

The action for the free field $S_{\mathrm{free},\phi}$ is invariant under the conformal transformation defined by
\begin{align}
 g_{\mu\nu}&\to \bar{g}_{\mu\nu}\coloneqq \left(C(x)\right)^2g_{\mu\nu}(x),\\
   \phi(x)\to\bar{\phi}(x) &\coloneqq (C(x))^{-\frac{(n+1)-2}{2}}\phi(x).
\end{align}
Now, let us impose a condition on the conformal factor such that
\begin{align}
C(x^\mu(\lambda))=\frac{\mathrm{d}\tau}{\mathrm{d}\lambda}=\frac{1}{\cos^2{\omega(\lambda)}+\left(\frac{\Omega}{\omega}\right)^2\sin^2{(\omega\lambda)}}\label{eq_constraint_C}
\end{align} 
is satisfied on the trajectory. Introducing a new trajectory $\bar{x}^\mu(\tau)\coloneqq x^{\mu} (\lambda(\tau))$ in the conformally related spacetime $(M,\bar{g}_{\mu\nu})$, $\tau$ is a proper time of the new trajectory in $(M,\bar{g}_{\mu\nu})$ since
\begin{align}
 \bar{g}_{\mu\nu}\frac{\mathrm{d}\bar{x}^\mu}{\mathrm{d}\tau}\frac{\mathrm{d}\bar{x}^\nu}{\mathrm{d}\tau}=C^2\left(\frac{\mathrm{d}\lambda}{\mathrm{d}\tau}\right)^2 g_{\mu\nu}\frac{\mathrm{d}x^\mu}{\mathrm{d}\lambda}\frac{\mathrm{d}x^\nu}{\mathrm{d}\lambda}=-1
\end{align}
holds.

Let us now consider a new  detector in $(M,\bar{g}_{\mu\nu})$ whose switching function $\chi_{\Omega}(\tau)$ satisfies
\begin{align}
  \chi_{\Omega}(\tau)=  \chi_{\omega}(\lambda(\tau)) C(x^\mu(\lambda(\tau))^{\frac{n-4}{2}}\label{eq_functions_consistency}.
\end{align}
Then, the action $S_{\mathrm{UDW}}$ can be rewritten as
\begin{align}
S_{\mathrm{UDW}}&= -\frac{1}{2}\int_{M} \mathrm{d}^{n+1} x \sqrt{-\bar{g}}\left(\bar{g}^{\mu\nu}\bar{\nabla}_{\mu}\bar{\phi}(x)\bar{\nabla}_\nu\bar{\phi}(x)\right.\nonumber\\
&\left.\quad\quad\quad\quad\quad\quad\quad\quad\quad\quad\quad\quad\quad\quad+\xi \bar{R}(x)\bar{\phi}(x)^2\right)\nonumber\\
&\quad+ \frac{1}{2}\int \mathrm{d}\tau \left(\left(\frac{\mathrm{d}q_{\Omega}(\tau)}{\mathrm{d}\tau}\right)^2-\Omega^2 q_{\Omega}^2(\tau)\right)\nonumber\\
&\quad -  c \int  \mathrm{d}\tau\sqrt{2\omega}\chi_{\Omega}(\tau)q_{\Omega}(\tau)\bar{\phi}(\bar{x}^{\mu}(\tau)),
\end{align}
where $\bar{\nabla}_\mu$ is the covariant derivative associated with $\bar{g}_{\mu\nu}$ and $\bar{R}(x)$ is the Ricci scalar defined by the conformally related metric $\bar{g}_{\mu\nu}$.
This shows the duality in dynamics among detectors in different curved spacetimes. 

It should be noted that we can take an arbitrary value for the parameter $\Omega$. Thus, there are infinite kinds of UDW detector systems in different conformally related spacetimes whose dynamics is described by the same action. 


So far, we have seen the duality in the classical action. The nonperturvative duality also holds in the unitary time-evolution operator for a quantized system since Eq. \eqref{eq_takagi_q} holds. 
In the following subsection, we will construct an example of conformal factor $C$ satisfying Eq.\eqref{eq_constraint_C} for a detector at rest.

\subsection{Example: Static detectors and spacetimes related by the conformal-Takagi transformation}
As an example, let us consider a static UDW detector with angular frequency $\omega$ in $(n+1)$-dimensional flat spacetime. 
The trajectory for this detector is given by
\begin{align}
 x^\mu(\lambda)=(\lambda,0,\ldots,0),
\end{align}
where $\lambda$ is the proper time. The constraint on the conformal factor is given by
\begin{align}
 C(\lambda,0,\ldots,0)=\frac{1}{\cos^2{(\omega\lambda)}+\left(\frac{\Omega}{\omega}\right)^2\sin^2{(\omega\lambda)}}.
\end{align}
A simple solution is 
\begin{align}
 C(t,x^i)=\frac{1}{\cos^2{(\omega t)}+\left(\frac{\Omega}{\omega}\right)^2\sin^2{(\omega t)}}.
\end{align}

The line element of the new spacetime is given by
\begin{align}
&\bar{\mathrm{d}s}^2=\bar{g}_{\mu\nu}\mathrm{d}x^\mu \mathrm{d}x^\nu\nonumber\\
 &=\frac{1}{\left(\cos^2{(\omega t)}+\left(\frac{\Omega}{\omega}\right)^2\sin^2{(\omega t)}\right)^2}\left(-\mathrm{d}t^2+\sum_{i=1}^n\left(\mathrm{d}x^i\right)^2\right).\label{eq_line_element_tfmd}
\end{align}
Introducing a new time variable \mbox{$T\coloneqq \frac{1}{\Omega}\arctan{\left(\frac{\Omega}{\omega}\tan{(\omega t)}\right)}$}, it can be recast into
\begin{align}
 \bar{\mathrm{d}s}^2=-\mathrm{d}T^2 +a(T)^2\sum_{i=1}^n\left(\mathrm{d}x^i\right)^2,
\end{align}
where the scale factor is defined by
\begin{align}
a(T) \coloneqq\cos^2{(\Omega T)}+\left(\frac{\omega}{\Omega}\right)^2\sin^2{(\Omega T)}.
\end{align}
The trajectory of the detector in the new spacetime $(M,\bar{g}_{\mu\nu})$ is also static and given by $(T,0,\ldots,0)$ in this coordinate system. 

It should be noted that the duality in the dynamics is applicable for any value of $\Omega$. For example, $\Omega\to 0$ corresponds to the power law inflation universe:
\begin{align}
\bar{\mathrm{d}s}^2=-\mathrm{d}T^2 +(1+\omega^2T^2) \sum_{i=1}^n\left(\mathrm{d}x^i\right)^2.\label{eq_line_element_plu}
\end{align}
In the case of $\Omega\to0$, it should be noted that $\lambda\in(-\frac{\pi}{2\omega},\frac{\pi}{2\omega})$ corresponds to $\tau\in(-\infty,\infty)$ in Eq.\eqref{eq_Takagi1}. This means that when we use the duality of dynamics for $\Omega=0$, we have to switch off the coupling between the detector and the field except for $\lambda\in(-\frac{\pi}{2\omega},\frac{\pi}{2\omega})$. 

This formalism can be extended to smeared detectors. Smeared detectors present additional complications coming from the fact that the different detector interaction and free Hamiltonians generate translations with respect to different proper times and Fermi-Walker frame transformations need to be considered when analyzing several smeared detectors in curved spacetimes (see \cite{martinmartinez2020general}). However, the Takagi transformation formalism would still be applicable in these cases. 

\section{Entanglement harvesting and negativity}\label{sec_negativity}
In the previous section, we have shown the duality in dynamics for a single harmonic oscillator UDW detector. The result can be straightforwardly extended to the multiple detectors if the conformal factor satisfies Eq.\eqref{eq_constraint_C} for all detector trajectories. 

An interesting application will be the entanglement harvesting, i.e., entanglement extraction from a field by using two UDW detectors. Consider two-UDW detectors in a spacetime $(M,g_{\mu\nu})$ coupled to the scalar field. Even if the detectors are spatially separated, they can be entangled after the interaction. The entanglement between detectors can be quantified negativity. For example, we can adopt the nonperturbative method developed in \cite{Brown2012,Bruschi:2012rx} for numerical calculation. The duality proven in the previous section implies that the state after the interaction can be also interpreted as the state for detectors coupled to a scalar field in $(M,\bar{g}_{\mu\nu})$. Therefore, one calculation on an entanglement harvesting protocol in $(M,g_{\mu\nu})$ can be mapped into the one in different spacetimes $(M,\bar{g}_{\mu\nu})$ parametrized by $\Omega$. It should be noted that the interpretation of the initial states of detectors will be different in each setup since they have different angular frequencies. For example, the vacuum state of a harmonic oscillator with frequency $\omega$ is mapped to a squeezed state of a Takagi-transformed oscillator of frequency $\Omega$.

In this section, we further show that leading-order calculations on negativity in entanglement harvesting using qubit detectors can be directly mapped to those with harmonic oscillators if detectors are initially in the ground states. Combining the duality in the dynamics, we can translate the results in prior researches for qubit detectors in a spacetime $(M,g_{\mu\nu})$ into the leading-order analyses for harmonic oscillator detectors in different spacetimes $(M,\bar{g}_{\mu\nu})$. The notations for this perturbative correspondence is summarized in Table \ref{table_notations}. 
The trajectory and the proper time are given, respectively, by
\begin{align}
 \tau_d^{\Omega_d}(\lambda_d)&\coloneqq \frac{1}{\Omega_d}\arctan{\left(\frac{\Omega_d}{\omega_d}\tan{(\omega_\mathrm{d}\lambda_d)}\right)},\\
 \bar{x}_d^\mu(\tau_d^{\Omega_d})&\coloneqq x_d^\mu(\lambda_d) .
\end{align}
The switching function in new spacetime $(M,\bar{g}_{\mu\nu})$ must satisfy
\begin{align}
 \chi_d^{\Omega}(\tau_d)&=\chi_d(\lambda_d(\tau_d))C^{\frac{n-4}{2}}(x^\mu(\lambda_d(\tau_d))).
\end{align}

\begin{table*}[t]
 \begin{tabular}{|c||c|c|c|}\hline
  &Qubit detector in $(M,g_{\mu\nu})$ &HO detector in $(M,g_{\mu\nu})$ & HO detector in $(M,\bar{g}_{\mu\nu})$\\\hline\hline
  Energy gap& $\omega_d$&--- &--- \\ \hline
  Angular frequency&--- &$\omega_d$ &$\Omega_d$: free parameter \\ \hline
  Coupling constant &$c_d$ &$c_d$ &$c_d$ \\ \hline
 Trajectory & $x_d^\mu(\lambda_d)$& $x_d^\mu(\lambda_d)$&$\bar{x}_d^\mu(\tau_d^{\Omega_d})$ \\ \hline
  Proper time& $\lambda_d$& $\lambda_d$& $\tau_d^{\Omega_d}(\lambda_d)$\\ \hline
  Switching function& $\chi_d(\lambda_d)$& $\chi_d(\lambda_d)$ & $\chi_d^{\Omega_d}(\tau_d)$\\ \hline
  Initial state&$\ket{g_d}$: ground state &$\ket{0_d}$: ground state & $\ket{0_d}$: squeezed state\\ \hline
 \end{tabular}
\caption{Summary of the notations.}\label{table_notations}
\end{table*}

\subsection{Qubit UDW detectors}
Consider qubit UDW detectors $A$ and $B$ with energy gap $\omega_d$ ($d=A,B$), coupled to the scalar field $\hat{\phi}(x)$. Let $x_d^\mu(\lambda)=(x_d^0(\lambda),\bm{x}(\lambda))$ be the trajectory of the detector in $(M,g_{\mu\nu})$, where $\lambda_d$ is the proper time. Let $(t,\bm{x})$ be the quantization frame of the field. 
In the interaction picture, the interaction Hamiltonian generating the time evolution with respect to $t$ for the qubit detector $d$ is given by
\begin{align}
 \hat{H}_{\mathrm{I,qubit}}^d(t)=c_d\gamma_d^{-1}(\lambda_d(t)) \chi_d(\lambda_d(t))\hat{\mu}_d(\lambda_d(t))\hat{\phi}(t,\bm{x}_d(t)),
\end{align}
where $c_d$ is the coupling constant, $\gamma_d$ is defined as $\gamma_d\coloneqq \frac{\mathrm{d}t}{\mathrm{d}\lambda_d}$, $\chi_d(\lambda)$ is the switching function and $\bm{x}_d(t)$ is the position of detector $d$ at $t$. The operator $\hat{\mu}_d(\lambda_d)$ is the monopole moment
\begin{align}
 \hat{\mu}_d(\lambda_d)=e^{-i\omega_\mathrm{d}\lambda_d}\hat{\sigma}_d^-+e^{i\omega_\mathrm{d}\lambda_d}\hat{\sigma}_d^+
\end{align}
describing the internal degrees of freedom of the detector. By using the ground state $\ket{g_d}$ and excited state $\ket{e_d}$, the operators are expressed by $\hat{\sigma}_d^-=\ket{g_d}\bra{e_d}$ and $\hat{\sigma}_d^+=\ket{e_d}\bra{g_d}$. 

The time-evolution unitary operator of the total system is given by
\begin{align}
  \hat{U}_{\mathrm{qubit}}=\mathcal{T}\exp{\left(-i\int_{-\infty}^\infty \mathrm{d}t \hat{H}_{\mathrm{I,qubit}}(t)\right)},
\end{align}
where $\hat{H}_{\mathrm{I,qubit}}\coloneqq \sum_{d=A,B}\hat{H}_{\mathrm{I,qubit}}^{d}$ and $\mathcal{T}$ denotes the time ordering. 
As is the case for most studies in quantum field theory, it is difficult to calculate the time evolution nonperturbatively for general switching functions. Let us expand the unitary operator as the Dyson series:
\begin{align}
 \hat{U}_{\mathrm{qubit}}=\mathbb{I}+\hat{U}^{(1)}+\hat{U}^{(2)}+\mathcal{O}(c_\mu^3),
\end{align}
where
\begin{align}
 \hat{U}^{(1)}&\coloneqq -i\int_{-\infty}^\infty \mathrm{d}t\hat{H}_{\mathrm{I,qubit}}(t),\\
 \hat{U}^{(2)}&\coloneqq (-i)^2\int_{-\infty}^\infty \mathrm{d}t \int_{-\infty}^t \mathrm{d}t' \hat{H}_{\mathrm{I,qubit}}(t)\hat{H}_{\mathrm{I,qubit}}(t').
\end{align}
Since we are interested in the entanglement extracted by the interaction, let us assume that the initial state of the total system is given by a product state:
\begin{align}
 \hat{\rho}_0=\hat{\rho}_{AB,0}\otimes \hat{\rho}_{\phi}
\end{align}
We further assume that the one-point function $\mathrm{Tr}(\hat{\rho}_\phi\hat{\phi}(x))$ vanishes. Physically important states such as the vacuum state, squeezed states or thermal states satisfy this condition. Then, the reduced state for the detector system is given by
\begin{align}
 \hat{\rho}_{AB}&=\hat{\rho}_{AB,0}+\hat{\rho}_{AB}^{(2,0)}+\hat{\rho}_{AB}^{(0,2)}+\hat{\rho}_{AB}^{(1,1)}+\mathcal{O}(c_{d}^4),
\end{align}
where we have defined
\begin{align}
 \hat{\rho}_{AB}^{(i,j)}\coloneqq \mathrm{Tr}_{\phi}\left(\hat{U}^{(i)}\hat{\rho}_{0}\left(\hat{U}^{(j)}\right)^\dag\right).
\end{align}

Let us assume that the detectors are initially in the ground state:
\begin{align}
 \hat{\rho}_{AB,0}=\ket{g_A}\bra{g_A}\otimes \ket{g_B}\bra{g_B}.
\end{align}
To describe $\hat{\rho}_{AB}$, it is convenient to adopt a matrix representation with respect to a basis
\begin{align}
\begin{split}
\ket{g_A,g_A}&=(1,0,0,0)^\dag,\\
\ket{e_A,g_B}&=(0,1,0,0)^\dag,\\
\ket{g_A,e_B}&=(0,0,1,0)^\dag,\\
\ket{e_A,e_B}&=(0,0,0,1)^\dag.
\end{split}
\end{align} 
In the leading-order calculation, the state is expressed as
\begin{align}
 \hat{\rho}_{AB}=
\begin{pmatrix}
 1-\mathcal{L}_{AA}-\mathcal{L}_{BB}&0&0&\mathcal{M}^*\\
 0&\mathcal{L}_{AA} &\mathcal{L}_{AB}&0\\
 0& \mathcal{L}_{AB}^*& \mathcal{L}_{BB}&0\\
 \mathcal{M}&0 & 0&0
\end{pmatrix}
+\mathcal{O}(c_d^4),\label{eq_rho_qubit}
\end{align}
where we have defined
\begin{align}
 \mathcal{L}_{dd'}&\coloneqq c_d c_{d'}\int_{-\infty}^\infty \mathrm{d}t \int_{-\infty}^\infty \mathrm{d}t'\nonumber\\
&\quad\quad\quad L_{d}(t)L_{d'}^*(t')W(t',\bm{x}_{d'}'(t'),t,\bm{x}_d(t))\label{eq_def_Ldd}\\
 \mathcal{M}&\coloneqq  -c_Ac_B \int_{-\infty}^\infty \mathrm{d}t \int_{-\infty}^t \mathrm{d}t' \nonumber\\ &\quad\left(L_{A}(t)L_{B}(t')W(t,\bm{x}_A(t),t',\bm{x}_B(t'))\right.\nonumber\\
 &\quad\quad\left.+L_{B}(t)L_{A}(t')W(t,\bm{x}_B(t),t',\bm{x}_A(t')\right),\label{eq_def_M}
\end{align}
and
\begin{align}
 L_d(t)\coloneqq \gamma_d^{-1}(\lambda_d(t))\chi_{d}(\lambda_d(t))e^{i\omega_\mathrm{d}\lambda_d(t)}.
\end{align}
Here, $W(x,x')\coloneqq \mathrm{Tr}(\hat{\rho}_{\phi} \hat{\phi}(x)\hat{\phi}(x'))$ denotes the Wightman function of the field. 

The negativity between two UDW detectors is defined as the sum of the negative eigenvalue for the partial transpose of $\hat{\rho}_{AB}$. In the leading order, the eigenvalue which can take a negative value is unique and given by
\begin{align}
 E_1&=\frac{1}{2}\left(\mathcal{L}_{AA}+\mathcal{L}_{BB}-\sqrt{\left(\mathcal{L}_{AA}-\mathcal{L}_{BB}\right)^2+4|\mathcal{M}|^2}\right)\nonumber\\
 &\quad+\mathcal{O}(c_d^4).\label{eq_def_E1}
\end{align}
From Eq.\eqref{eq_rho_qubit}, there is another eigenvalue
\begin{align}
 E_2=-|\mathcal{L}_{AB}|^2.
\end{align}
However, this term is $\mathcal{O}(c_d^4)$. Although we need to calculate the higher-order contribution for $\hat{\rho}_{A B}$ in order to obtain its correct form, $E_2$ does not contribute to the leading order of negativity. For a related argument on this point, see \cite{Pozas-Kerstjens:2015}. Neglecting the higher-order contributions, we get the negativity in the leading order as
\begin{align}
 \mathcal{N}_{\mathrm{qubit}}=\max\{-E_1,0\}+\mathcal{O}(c_d^4).\label{eq_neg_qubit}
\end{align}

\subsection{Harmonic oscillator UDW detectors}
Let us perform the same calculation for UDW detectors $A$ and $B$ whose internal degrees of freedom are described by harmonic oscillator with angular frequency $\omega_A$ and $\omega_B$, respectively. We assume that both detectors have unit mass. We adopt the same notation for the coupling constant, the trajectory and switching functions as in the previous subsection. 

The interaction Hamiltonian is given by
\begin{align}
 \hat{H}_{\mathrm{I,HO}}^d(t)
&=c_d\gamma_d^{-1}(\lambda_d(t)) \chi_d(\lambda_d(t))\sqrt{2\omega_d}\hat{q}_d(\lambda_d(t))\hat{\phi}(t,\bm{x}_d(t)),
\end{align}
where we have defined
\begin{align}
 \hat{q}_d(\lambda)&\coloneqq e^{i\hat{H}_{\omega_d}\lambda}\hat{q}e^{-i\hat{H}_{\omega_d}\lambda},
\end{align}
and $\hat{H}_d\coloneqq \frac{\omega_d^2}{2}\hat{q}_d^2+\frac{1}{2}\hat{p}_d^2$. The factor $\sqrt{2\omega_d}$ is included so that $\hat{H}_{\mathrm{I,HO}}$ has correct dimension. Introducing the creation and annihilation operators $\hat{a}_d^\dag$ and $\hat{a}_d$ as
\begin{align}
 \hat{a}_d\coloneqq \frac{1}{\sqrt{2}}\left(\sqrt{\omega_d}\hat{q}_d+i\frac{1}{\sqrt{\omega_d}}\hat{p}_d\right),
\end{align}
we get
\begin{align} 
\hat{q}_d(\lambda)
&=\frac{1}{\sqrt{2\omega_d}}\left(e^{-i\omega_\mathrm{d}\lambda}\hat{a}_d+e^{i\omega_\mathrm{d}\lambda}\hat{a}_d^\dag\right).
\end{align}
The ground state $\ket{0}_d$ for the Hamiltonian $\hat{H}_d$ is given by the following condition:
\begin{align}
 \hat{a}_d\ket{0_d}=0.
\end{align}

Let us assume that the field is in a state with a vanishing one-point function and the detectors are in the ground state:
\begin{align}
 \hat{\rho}_0=\ket{0_A}\bra{0_A}\otimes\ket{0_B}\bra{0_B}\otimes \hat{\rho}_{\phi}.
\end{align}
The unitary time-evolution operator is given by
\begin{align}
 \hat{U}_{\mathrm{HO}}\coloneqq \mathcal{T}\exp{\left(-i\int_{-\infty}^\infty \mathrm{d}t \hat{H}_{\mathrm{I,HO}}(t)\right)},\label{eq_unitary_HO}
\end{align}
where $\hat{H}_{\mathrm{I,HO}}\coloneqq \sum_{d=A,B}\hat{H}_{\mathrm{I,HO}}^d$.
As we have done in the previous section, we can calculate $\hat{\rho}_{AB}=\mathrm{Tr}_{\rho}\left(\hat{U}_{\mathrm{HO}}\rho_{0}\hat{U}_{\mathrm{HO}}^\dag\right)$  as a series. In the case of the harmonic oscillator, the second excited state also contributes to the leading-order calculation. Fixing vectors
\begin{align}
\begin{split}
 \ket{0_A,0_B}&=(1,0,0,0,0,0)^\dag,\\ \ket{1_A,0_B}&=(0,1,0,0,0,0)^\dag,\\ \ket{0_A,1_B}&=(0,0,1,0,0,0)^\dag,\\
\ket{1_A,1_B}&=(0,0,0,1,0,0)^\dag,\\
 \ket{2_A,0_B}&=(0,0,0,0,1,0)^\dag,\\ \ket{0_A,2_B}&=(0,0,0,0,0,1)^\dag,
\end{split}
\end{align}
the state $\hat{\rho}_{AB}$ is written (to second order in $c_d$) as
\begin{align}
 \hat{\rho}_{AB}=
\begin{pmatrix}
 1-\mathcal{L}_{AA}-\mathcal{L}_{BB}&0&0&\mathcal{M}^*&\mathcal{N}_A^*&\mathcal{N}_B^*\\
 0& \mathcal{L}_{AA}& \mathcal{L}_{AB}&0 &0 &0 \\
 0& \mathcal{L}_{AB}^*& \mathcal{L}_{BB}& 0&0 &0 \\
 \mathcal{M}&0 &0 & 0& 0&0\\
\mathcal{N}_A&0 &0 & 0& 0&0\\
\mathcal{N}_B&0 &0 & 0& 0&0 
\end{pmatrix},
\label{eq_rho_HO}
\end{align}
where $\mathcal{L}_{dd'}$ and $\mathcal{M}$ are defined in Eqs.\eqref{eq_def_Ldd} and \eqref{eq_def_M} and
\begin{align}
 \mathcal{N}_d&\coloneqq- \sqrt{2}c_d^2 \int_{-\infty}^\infty \mathrm{d}t\int_{-\infty}^t \mathrm{d}t'\nonumber\\
 &\quad\quad\quad\quad L_d(t)L_d(t')W(t,\bm{x}_d(t),t',\bm{x}_d(t')).
\end{align}
The difference between the qubit detectors and harmonic oscillator detectors only appears in off-diagonal term $\mathcal{N}_d$. This fact has been pointed out in \cite{Brown2012}. 
From this expression, we can calculate the negativity. A potentially negative eigenvalue in the leading-order calculation is again given by $E_1$ defined in \eqref{eq_def_E1}. There is another eigenvalue $E_2'=-(|\mathcal{L}_{AB}|^2+|\mathcal{N}_A|^2+|\mathcal{N}_B|^2)$ calculated from Eq.\eqref{eq_rho_HO}, but this is not the leading-order contribution: $E_2'=\mathcal{O}(c_d^4)$. Therefore, the negativity is given by
\begin{align}
 \mathcal{N}_{\mathrm{HO}}=\max\{-E_1,0\}+\mathcal{O}(c_d^4).\label{eq_neg_ho}
\end{align}
It implies that the difference between the harmonic oscillator and qubit detectors only appears in the higher-order corrections.

\subsection{Example}
\begin{figure}[h]
\centering
 \includegraphics[height=4cm]{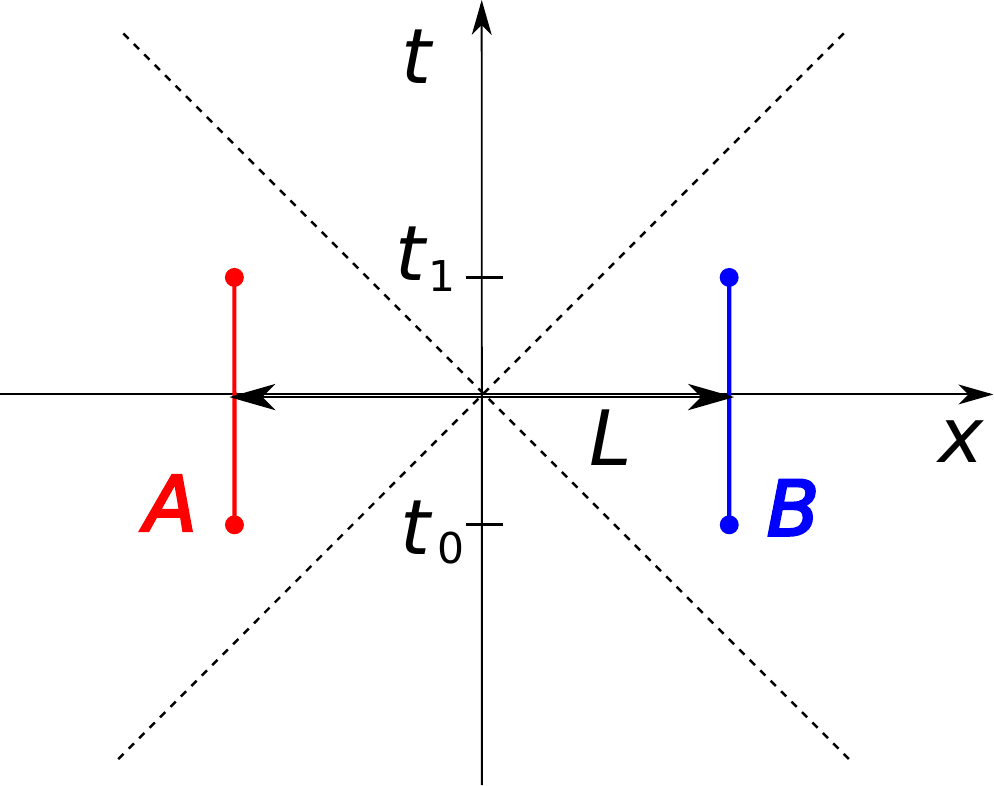}
\caption{\label{fig_eh} A schematic figure of entanglement harvesting using two pointlike static UDW detectors with spatial separation $L$. The detectors $A$ and $B$ couple to the scalar field for a finite time duration $t_0<t<t_1$. It is assumed that $t_1-t_0<L$ holds so that two detectors are causally disconnected.}
\end{figure}

We here provide a concrete example where our results established so far are applicable. Consider two pointlike qubit UDW detectors $A$ and $B$ which couple to the massless scalar field in the $(n+1)$-dimensional Minkowski spacetime whose line element is given by
\begin{align}
    \mathrm{d}s^2=-\mathrm{d}t^2 +\sum_{i=1}^n (\mathrm{d}x^{i})^2.
\end{align}
We assume that the detectors and the field are initially in their ground state and that the switching functions $\chi_A(t)$ and $\chi_B(t)$ vanish unless $t_{0}< t <t_1$. Let $L$ be the spatial separation of the detectors. If $t_1-t_0< L$, the detectors are causally disconnected. Therefore, if detectors get entangled after the interaction, it implies that they extract spatial entanglement from the scalar field. See Fig. \ref{fig_eh} for a schematic picture of the setup.

The negativity between the detectors $A,B$ is commonly used to diagnose the entanglement. 
For example, in \cite{Reznik2003}, the author investigated the negativity with $n=3$ for
\begin{align}
    \chi_A(t)=\chi_B(t)=
    \begin{cases}
    \cos^2(\pi t) \quad &(\text{if } t_0 < t <t_1)\\
    0 \quad &(\text{otherwise})
    \end{cases},
\end{align}
where $t_0$ and $t_1$ are fixed as $-t_0=t_1=1/2$. It is shown that the leading-order negativity is positive with particular energy gaps $\omega_A$ and $\omega_B$ even if $t_1-t_0<L$. Although the negativity itself is not explicitly evaluated in \cite{Reznik2003}, it can be done numerically.  

\begin{figure}[h]
\centering
 \includegraphics[height=4cm]{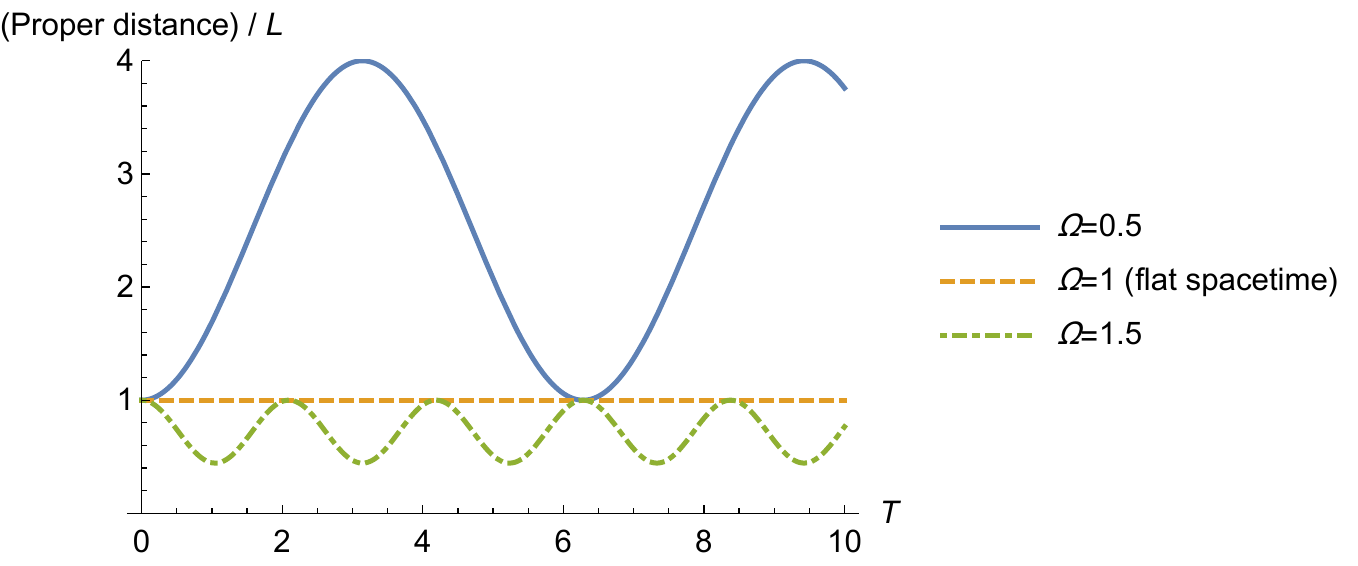}
\caption{\label{fig_pd} The plot for the proper distance between the detectors scaled by $L$. The angular frequency $\omega$ of the harmonic oscillators in the flat spacetime is set to unity. The conformally flat spacetime related via the conformal-Takagi transformation is parameterized by $\Omega$ whose line element is given by Eq.\eqref{eq_le}. }
\end{figure}

Now, let us apply our results to this setup. Equations \eqref{eq_neg_qubit} and \eqref{eq_neg_ho} imply that the negativity of qubit detectors with energy gaps $\omega_d$ is equivalent to that of harmonic oscillators with angular frequency $\omega_d$ in the same flat spacetime in the leading-order calculation. By using the conformal-Takagi transformation, it is also equivalent to the leading-order negativity between detectors extracted from the scalar field in curved spacetimes. Especially when the detectors' frequencies are the same, i.e., $\omega\coloneqq \omega_A=\omega_B$ and $\Omega\coloneqq \Omega_A=\Omega_B$,  a metric for the spacetime related by the conformal-Takagi transformation is given by
\begin{align}
    &\bar{\mathrm{d}s}^2=-\mathrm{d}T^2+a(T)^2\sum_{i=1}^n\left(\mathrm{d}x^i\right)^2,\label{eq_le}
\end{align}
where we have defined
\begin{align}
    T&\coloneqq \frac{1}{\Omega}\arctan\left(\frac{\Omega}{\omega}\tan\left(\omega t\right)\right),\\
    a(T)&\coloneqq \cos^2\left(\Omega T\right)+\left(\frac{\omega}{\Omega}\right)^2 \sin^2\left(\Omega T\right).
\end{align}
In this spacetime, the detectors are at rest and their comoving distance is $L$. Since the scale factor depends on $T$, the proper distance between the detectors changes in time. 
In Fig. \ref{fig_pd}, the time dependence of the proper distance is plotted. Depending on $\Omega$, the proper distance oscillates in different ways.


\section{Conclusion}
Obtaining analytically or numerically manageable results for spatially smeared particle detectors interacting for finite times following arbitrary trajectories in curved spacetimes is a very challenging problem. We have shown that one can apply the Takagi transformations to map difficult problems involving Unruh-DeWitt detectors in curved spacetimes to simpler problems in conformally related spacetimes. Hence, the tools developed in this paper open the door to the analytical and numerical study of new scenarios of relativistic quantum information and of quantum field theory in curved spacetimes. As an example we have shown the case of a simple mapping that relates simple stationary detectors in flat spacetime to detectors in cosmological scenarios. Furthermore we have shown how to apply these techniques to entanglement harvesting and discussed how they can be applied to both harmonic oscillator detectors as well as qubit detectors by showing that the leading-order results are equivalent in both cases. 

This work paves the way to further studies on entanglement harvesting in scenarios that were previously analytically inaccessible. Future explorations will include the study of smeared detectors in conformally flat scenarios, and the study of the effect of the dynamics of the center of mass of the detector in the harvesting of spacelike entanglement, following up on recent results that extend particle detector models to account for the dynamics of their center of mass degree of freedom \cite{Nadine}. 

\begin{acknowledgments}
The authors would like to thank N. C. Menicucci for valuable discussion. This research had started at the 12th Relativistic Quantum Information Workshop in Brisbane. We would like to thank the hospitality of the organizers. This research was partially supported by JSPS KAKENHI Grants No. 18J20057 (K.Y.) and No. JP19K03838 (M.H.), and by Graduate Program on Physics for the Universe of Tohoku University (K.Y. and T.T.). E.M.-M. and A. K. are partially supported by the NSERC Discovery Program. E. M.-M. also acknowledges funding of his Ontario Early Researcher Award. A.K. is also partially supported by a Google Faculty Research Award. 
\end{acknowledgments}

\bibliography{references}
\clearpage
\widetext
\appendix
\section*{Appendix: The Takagi transformation}
Here we derive Eq.\eqref{eq_takagi_q}. 
Let us consider a product of unitary operators $\hat{W}_{f_\omega,g_\omega}(\lambda)\hat{U}_{\omega}(\lambda)$, defined by
\begin{align}
 \hat{W}_{f_\omega,g_\omega}(\lambda)\coloneqq e^{-i\frac{f_\omega(\lambda)}{2}\left(\hat{q}\hat{p}+\hat{p}\hat{q}\right)}e^{-i\frac{g_\omega(\lambda)}{2}\hat{q}^2},\quad \hat{U}_{\omega}(\lambda)\coloneqq e^{-i\left(\frac{\omega^2}{2}\hat{q}^2+\frac{1}{2}\hat{p}^2\right)\lambda}
\end{align}
for real functions $f_\omega$ and $g_\omega$ satisfying $f_\omega(0)=g_\omega(0)=0$. Its derivative with respect to $\lambda$ is given by
\begin{align}
& i\frac{d}{\mathrm{d}\lambda}\left(\hat{W}_{f_\omega,g_\omega}(\lambda)\hat{U}_\omega(\lambda)\right)\nonumber\\
&=\left(\left(i\frac{d}{\mathrm{d}\lambda}\hat{W}_{f_\omega,g_\omega}(\lambda)\right)\hat{W}_{f_\omega,g_\omega}(\lambda)^\dag+\hat{W}_{f_\omega,g_\omega}(\lambda)\left(\frac{\omega^2}{2}\hat{q}^2+\frac{1}{2}\hat{p}^2\right)\hat{W}_{f_\omega,g_\omega}(\lambda)^\dag\right)\hat{W}_{f_\omega,g_\omega}(\lambda)\hat{U}_\omega(\lambda).\label{eq_sch}
\end{align}

Since the identity
\begin{align}
&\left(i\frac{d}{\mathrm{d}\lambda}\hat{W}_{f_\omega,g_\omega}(\lambda)\right)\hat{W}_{f_\omega,g_\omega}(\lambda)^\dag+\hat{W}_{f_\omega,g_\omega}(\lambda)\left(\frac{\omega^2}{2}\hat{q}^2+\frac{1}{2}\hat{p}^2\right)\hat{W}_{f_\omega,g_\omega}(\lambda)^\dag\nonumber\\
  &=\frac{f_\omega'(\lambda)}{2}\left(\hat{p}\hat{q}+\hat{q}\hat{p}\right)+\frac{g_\omega'(\lambda)}{2}e^{2f_\omega(\lambda)}\hat{q}^2+\frac{\omega^2}{2}e^{2f_{\omega}(\lambda)}\hat{q}^2+\frac{1}{2}\left(e^{-f_{\omega}(\lambda)}\hat{p}+g_{\omega}(\lambda)e^{f_{\omega}(\lambda)}\hat{q}\right)^2\nonumber\\
 &=\frac{ e^{2f_\omega(\lambda)}}{2}\left(g_\omega'(\lambda)+\omega^2+g_\omega^2 (\lambda)\right)\hat{q}^2+\frac{1}{2}\left(f_\omega'(\lambda)+g_\omega(\lambda)\right)\left(\hat{p}\hat{q}+\hat{q}\hat{p}\right)+\frac{e^{-2f_{\omega}(\lambda)}}{2}\hat{p}^2
\end{align}
holds, if the functions satisfy
\begin{align}
 g_\omega'(\lambda)+\omega^2+g_\omega^2 (\lambda)=0\label{eq_g}
\end{align}
and
\begin{align}
 f_\omega'(\lambda)+g_\omega(\lambda)=0,\label{eq_f}
\end{align}
then Eq.\eqref{eq_sch} yields
\begin{align}
 i\frac{\mathrm{d}}{\mathrm{d}t}\left(\hat{W}_{f_\omega,g_\omega}(\lambda)\hat{U}_\omega(\lambda)\right)=\frac{\hat{p}^2}{2}\hat{W}_{f_\omega,g_\omega}(\lambda)\hat{U}_\omega(\lambda),
\end{align}
where we introduced a new parameter $T$ satisfying $\frac{\mathrm{d}t}{\mathrm{d}\lambda}=e^{2f_{\omega}(\lambda)}$.
The solution for Eqs.\eqref{eq_g} and \eqref{eq_f} is given by
\begin{align}
f_\omega(\lambda)=-\ln{\left(\cos{(\omega\lambda)}\right)} ,\quad g_{\omega}(\lambda)=-\omega\tan{(\omega\lambda)}
\end{align}
under the constraints $f_{\omega}(0)=g_{\omega}(0)=0$. Therefore,
\begin{align}
 \hat{V}_{\omega}(\lambda)\hat{U}_{\omega}(\lambda)=e^{-i\frac{1}{2}\hat{p}^2 T}
\end{align}
holds for $T=\frac{1}{\omega}\tan{(\omega\lambda)}$. Since this equation holds for an arbitrary $\omega$, we have proven Eq.\eqref{eq_takagi_q} where $\tau$ is defined by 
\begin{align}
 \tau=\frac{1}{\Omega}\arctan{(\Omega T)}=\frac{1}{\Omega}\arctan{\left(\frac{\Omega}{\omega}\tan{(\omega\lambda)}\right)}.
\end{align}

\end{document}